\documentclass[12pt]{article}
\usepackage{amsfonts}

\begin{document}

\def\ba{\begin{eqnarray}}
\def\ea{\end{eqnarray}}
\def\a{\alpha}
\def\b{\beta}
\def\k{\kappa}
\def\p{\partial}
\def\o{\omega}
\def\w{\wedge}
\def\f{\frac}
\def\d{\delta}
\def\hb{\hbar}
\def\g{\gamma}
\def\s{\sigma}
\def\ve{\varepsilon}
\def\th{\theta}
\def\vp{\varphi}
\def\pr{\prime}
\def\ua{\uparrow}
\def\da{\downarrow}
\def\Th{\Theta}
\def\D{\Delta}
\def\L{\Lambda}
\def\G{\Gamma}
\def\O{\Omega}

\begin{titlepage}
\title{ Neutrino Oscillations Induced by\\ Space-Time Torsion  }
\author{ M. Adak$^{1,2}$, T. Dereli$^{2}$,  L. H. Ryder$^{1}$\\
$^{1}${\small Department of Physics, University of Kent, Canterbury,
Kent CT2 7NF, UK}\\
$^{2}${\small Department of Physics, Middle East Technical University,
 06531 Ankara, Turkey  } }
\date{27 July 2000}
\maketitle

\begin{abstract}

The gravitational neutrino oscillation problem is studied by
considering the Dirac Hamiltonian in a Riemann-Cartan space-time
and calculating the dynamical phase. Torsion contributions which
depend on the spin direction of the mass eigenstates are found.
These effects are of the order of Planck scales.

\end{abstract}
\end{titlepage}

\section{Introduction}

Neutrinos are the only particles for which a poem has been
written; "Cosmic Gall" by John Updike, quoted in \cite{r1}.
Recently neutrinos have attracted a lot of attention in high
energy physics; see \cite{r1,r2} and references therein. A major
problem at present is the problem of solar neutrinos.
Thermonuclear reactions in the sun produce large number of
electron neutrinos; the standard solar model allows us to predict
the flux of these neutrinos approximately as $ 7.3 \pm 2.3 $
SNU\footnote{SNU stands for solar neutrino unit and  1 SNU = 1
capture/sec/$10^{36}$ target atoms  }. On the other hand
experiments which have been done for decades give a measured flux
of $ 2.55 \pm 0.17 \pm 0.18 $ SNU -- approximately one third of
the theoretical rate. This is the solar neutrino problem. A
much-publicised solution is "neutrino oscillation (or mixing)".
According to neutrino oscillation, in contrast to the conventional
view that neutrinos are massless particles, they actually have a
mass different from zero, although very small. The essential idea
belongs to B.Pontecorvo \cite{pon}. Technically speaking it is
that the neutrino states of definite mass do not coincide with the
weak interaction eigenstates. In other words, an electron neutrino
emitted by sun becomes a linear combination of all three neutrinos
(electron, muon and tau) while travelling in space towards the
earth. Therefore the probability of detecting $ \nu_e $ emitted by
sun as $ \nu_\mu $ or $ \nu_\tau $ on the earth is different from
zero. This situation, of course, invokes physics beyond the
standard model. Although there are three neutrino families, the
essential physics of neutrino oscillations is illustrated very
well by considering the interactions of only two of these, which
we choose to be  $ \nu_e $ and $ \nu_\mu $.

All the above arguments have been cast in  Minkowski spacetime.
However we know that we live in a curved spacetime -- perhaps even
in a curved spacetime with torsion. Therefore, in more recent
years, physicists have turned their attention to specifically
gravitational contributions to neutrino oscillations -- see
\cite{r5,r6,r7,r8,r9,r10} and references therein.

In this paper we reinvestigate the effects of gravitation, but we
consider a spacetime with torsion, as well as simply curvature;
this is the Einstein-Cartan theory, or more specifically the
Einstein-Cartan-Dirac theory in our case, since we will be
considering Dirac particles in the spacetime. Some work along
these lines has already been done \cite{r9,r10}, but our results
do not entirely replicate these findings. The essence of our work
is to calculate the dynamical phase of neutrinos, by finding the
form of the Hamiltonian, H, from the Dirac equation in
Riemann-Cartan spacetime. The phase then follows from the formula
 \ba
   i\hb \f{\p \psi }{\p t } = H \psi \;\; \Longrightarrow \;\;
             \psi (t) = e^{-\f{i}{\hb }\int{H dt} } \psi (0) \; .
 \ea
In what follows, $\psi $ is a Dirac spinor and $H$ is a $4\times
4$ matrix; the exponential function is therefore well-defined, as
usual, by the series expansion. The Hamiltonian $H$ will depend,
for example, on momentum $\vec{p}$, and this is expressed not as a
differential operator but simply as a vector.

After we have introduced briefly the Einstein-Cartan-Dirac theory
in Sec.2 we find the hamiltonian of a Dirac particle in
subsec.2.1. In sec.3 we investigate, in turn, the neutrino
oscillations for azimuthal and radial motions. The conclusions are
given in sec.4.

\section{Einstein-Cartan-Dirac Theory}

A Riemann-Cartan space-time
is defined by the triple
$ \{ M, g, \nabla \} $
where M is
a 4-dimensional differentiable manifold,
equipped with a Lorentzian metric $g$ and a metric compatible
connection $\nabla$.   The metric tensor may be written in terms of
orthonormal basis 1-forms
$ \{ e^a \} $ as
 \ba
g = \eta_{ab} e^a \otimes e^b &=& - e^0 \otimes e^0 +  e^1 \otimes e^1
   +  e^2 \otimes e^2 +   e^3 \otimes e^3  \; , \label{eq1}
 \ea
while the connection is specified by a set of connection 1-forms
$ \{ {\G^a}_b \} $.
Metric compatibility requires
 \ba
D\eta_{ab} \equiv d \eta_{ab} - {\G^c}_a \eta_{cb} -
                   {\G^c}_b \eta_{ac}  = 0 \; , \label{eq2}
 \ea
so that  $ \G_{ab} = - \G_{ba} $. The Cartan structure equations
below define the torsion 2-forms $ \{ T^a \} $ and the curvature
2-forms   $ \{ {R^a}_b \} $  of the space-time:
 \ba
  De^a &\equiv&  de^a + {\G^a}_b \w e^b  = T^a \; , \label{eq3} \\
   D{\G^a}_b &\equiv& d{\G^a}_b + {\G^a}_c \w {\G^c}_b  =  {R^a}_b
   \; . \label{eq4}
 \ea
In this study d, D, $\imath_a$, * denote the exterior derivative,
the covariant exterior derivative, the interior derivative and the
Hodge star operator, respectively. The local orthonormal frame $
\{ X_a \} $ is dual to the coframe $ \{ e^a \} $;
 \ba
        \imath_{X_a}e^b \equiv \imath_a e^b =
            e^b(X_a) =  \delta_a^b \; .
 \ea
 The space-time orientation is set by the choice $ \epsilon_{0123}
= +1 $. In addition, the Dirac matrices $ \g^a $ are in the
standard Bjorken and Drell form~\cite{bjo} but for our metric
convention they satisfy
 \ba
       \g^a \g^b + \g^b \g^a =-2\eta^{ab}1_{4 \times 4} \; .
 \ea

We use the formalism of Clifford algebra-valued exterior forms for the
Dirac particles. In this formalism
 $ \g = \g^a e_a $
and
 $ *\g = \g^a *e_a $.
Furthermore,  the covariant exterior derivative of spinor fields
in the Riemann-Cartan space-time is
 \ba
    D\psi = d\psi - \f{1}{8} \G_{ab} [ \g^a,\g^b]\psi \; . \label{eq5}
 \ea

The connection 1-forms, $ \{ {\G^a}_b \}, $ are decomposed as
follows
 \ba
     {\G^a}_b = {\o^a}_b +{K^a}_b \; . \label{eq9}
 \ea
Here, $ \{ {\o^a}_b \} $ are Levi-Civita connection 1-forms which
are given by
 \ba
    de^a + {\omega^a}_b \w e^b = 0 \label{eq10}
 \ea
and  $ \{ {K^a}_b \} $ are kontorsion 1-forms which are given by
 \ba
    {K^a}_b \w e^b = T^a \; . \label{eq11}
 \ea

The Einstein-Cartan-Dirac Lagrangian density 4-form is
 \ba
    L  = -\f{1}{2 \ell^2 } R_{ab} \w *(e^a \w e^b ) +
       \f{i}{2} \{ \overline{\psi} *\g \w D\psi +\overline{D\psi}
          \w *\g \psi \} + \f{*m c}{\hb} \overline{\psi} \psi
           \; . \label{eq6}
 \ea
 The field equations are then \cite{tru} , \cite{r3}
 \ba
  i\hb *\g \w (D \psi - \f{1}{2} T \psi) +
       *m c \psi &=&0 \; ,\label{eq7} \\
  R^{bc} \w *(e_a \w e_b \w e_c ) &=& i \ell^2  ( \overline{\psi}
  *\g \;  \imath_a D\psi - ( \imath_a \overline{D\psi} )
            *\g \psi ) \; , \label{cur} \\
  T^c \w *(e_a \w e_b \w e_c ) &=& \f{ \ell^2 }{2}
     e_a \w e_b \w \overline{\psi} \g \g_5 \psi \; , \label{tor}
 \ea
where $ T= \imath_a T^a $ is the torsion trace 1-form and $ \ell =
\sqrt{ \f{Gh}{c^3 } } \; \sim 10^{-31} \; \; m $ is the Planck
length. As is seen from eqn.(\ref{tor}) the source of torsion is
spin and magnitude of torsion is proportional to square of the
Planck length. In this work, the spinorial contributions to the
curvature will be assumed small, and therefore neglected. That is
both the Dirac stress-energy tensor on the right hand side of
eqn.(\ref{cur}) and the torsion dependent terms on the left hand
side are dropped.

The Dirac equation (\ref{eq7}) can be recast as
 \ba
  i\hb \g^a \imath_a (D\psi -
            \f{1}{2} T \psi )-mc\psi = 0 \; . \label{eq8}
 \ea
Here the torsion trace 1-form $T$ is cancelled by a similar term
in the covariant exterior derivative and what is left behind are
the totally anti-symmetric axial components of the torsion. These
are going to be treated as given background fields.

\subsection{Hamiltonian of a Dirac Particle}

As mentioned above, our aim is to calculate the phase of a
neutrino beam in Riemann-Cartan spacetime. We do this by finding
the Hamiltonian for a Dirac particle. The spacetime we consider is
Schwarzschild spacetime and the metric is therefore given by
 \ba
    g = -f^2 dt \otimes dt + \f{1}{f^2} dr \otimes dr +
         r^2 (d \theta \otimes d\theta +
        \sin^2 \theta d\varphi \otimes d \varphi ) \; . \label{eq12}
 \ea
The orthonormal basis coframe is (from (\ref{eq1}))
 \ba
    e^0 = f cdt \hskip 1cm e^1 = \f{1}{f} dr \hskip 1cm
  e^2 = rd\theta \hskip 1cm e^3 = r \sin\theta d\varphi \label{eq13}
 \ea
where $ f^2 = 1- \f{2MG }{rc^2 } $. Using (\ref{eq13}) and
(\ref{eq10}) we can calculate the Levi-Civita connection 1-forms,
$ {\omega^a}_b $. We then substitute eqn.(\ref{eq9}) and the
Levi-Civita 1-forms into eqn.(\ref{eq5}) and then into
eqn.(\ref{eq8}). Now we use the identity
 \ba
   \g^a [ \g^b , \g^c ] = -2 \eta^{ab} \g^c +  2 \eta^{ac} \g^b
        +  2i \epsilon^{abcd} \g_5 \g_d
 \ea
and eqn.(\ref{eq11}), and then write the contorsion 1-forms in component
form as
  $ {K^a}_b = {K^a}_{bc} e^c $.
The resulting Dirac equation is
 \ba
    \{ \f{i \hb }{cf} \g^0 \f{\partial}{\partial t} +
    i \hb f \g^1 ( \f{\partial }{\partial r} + \f{1}{r} +
       \f{ f^{\prime} }{2f} )
    + \f{i \hb}{r} \g^2 ( \f{ \partial }{ \partial \theta } +
    \f{ \cot\theta }{2} ) \nonumber \\
   +  \f{ i \hb }{r \sin\theta }\g^3 \f{\partial }{\partial \varphi} -
   mc + \f{\hb}{4} K_{abc} \epsilon^{abcd} \g_5 \g_d \} \psi = 0
    \label{eq14}
 \ea
where $ \g_5 = -i\g^0 \g^1 \g^2 \g^3 $. From the last term we see
that it is only the totally anti-symmetric part of the contorsion
tensor (in other words, the purely axial component) which
contributes to the Dirac equation. [Actually, in theories in which
spin is the source of torsion, the contorsion tensor is in any
case totally anti-symmetric \cite{ryd}.] Relative to the
orthonormal basis 1-forms $\{ e^a \} $, we may write
 \ba
   K_{abc} = \epsilon_{abcf} A^f \label{vec}
 \ea
for some vector field\footnote{ From eqn.(\ref{eq13}) $A_0=
\frac{1}{fc}A_t \; , \;\; A_1= f A_r \; , \;\; A_2 = \frac{1}{r}
A_\theta \; , \;\; A_3 = \frac{1}{r\sin\theta}A_\varphi $. This
means $\vec{A} = (A_1,A_2,A_3)$ are not the cartesian components,
rather the spherical components in the orthonormal basis. } $ A=
A^a X_a $. Noting that
 $ \epsilon_{abcf} \epsilon^{abcd} = -3! \d^d_f $
and defining  the components of the momentum operator by
 \ba
    p_r &=& - i \hb  ( \f{\partial }{\partial r} + \f{1}{r} +
       \f{ f^{\prime} }{2f} )  \\
    p_\th &=& - \f{i \hb}{r}  ( \f{ \partial }{ \partial \theta } +
    \f{ \cot\theta }{2} )  \\
   p_\vp &=& -  \f{ i \hb }{r \sin\theta } \f{\partial }{\partial
             \varphi} \; ,
 \ea
the eqn.(\ref{eq14}) becomes
 \ba
     i \hb  \g^0 \f{\partial \psi }{\partial t} &=&
    \{ f^2 \g^1 p_r c + f \g^2 p_\th c +  f \g^3 p_\vp c \nonumber \\
    &{ }& + fmc^2  + \f{3}{2} \hb c f \g_5\g_a A^a \}\psi \; .
 \ea
Using  $ \b \equiv \g^0 $,
  $ \vec{ \alpha} = \beta \vec{\g} $
and $\g^0 (\g_5\g_a)A^a = \g_5A^0 + \vec{\Sigma}.\vec{A}  $ where
$ A^a=(A^0,\vec{A}) $ and $ \vec{\Sigma} =  \left (
                 \begin{array}{cc}
                                   \vec{\sigma} &   0         \\
                                        0       & \vec{\sigma}
                 \end{array}
          \right )  $.
Comparing the above with the Schr\"odinger equation
 \ba
    i \hb \f{ \p \psi }{\p t } = H \psi
 \ea
we find that
 \ba
    H = f^2 \a^1 p_r c + f \a^2 p_\th c + f \a^3 p_\vp c + f \beta mc^2
          + \f{3}{2} \hb c f A^0 \g_5
          + \f{3}{2} \hb c f \vec{\Sigma}.\vec{A} \; . \label{eq15}
 \ea

\section{Neutrino Oscillations}

Neutrinos that are produced in the sun arrive on the earth.
According to the theory of neutrino oscillations, a neutrino with
fixed flavour which is emitted by sun becomes a superposition (in
our model) of two neutrinos, i.e., they mix during the journey. In
this case, even if only one kind of neutrino were emitted by the
sun, the probability of measuring the other kind of neutrino on
the earth would not be zero. This is called "neutrino oscillation"
and is measured by the interference pattern observed on the earth.
The starting point of neutrino oscillation theory is that  flavour
eigenstates are  different from mass eigenstates;  each flavour
eigenstate is a linear superposition of mass eigenstates;
 \ba
     \nu_e &=& \cos\Th \nu_1 + \sin\Th \nu_2 \label{mix}  \\
     \nu_\mu &=& -\sin\Th \nu_1 + \cos\Th \nu_2 \label{mix1}
 \ea
where $\nu_e $ and $ \nu_\mu $ denotes electron and muon states,
respectively, and $ \nu_1 $ and $ \nu_2 $ mass eigenstates, and $
\Th $ is a mixing angle. It should be clear that this scheme only
works if the neutrino masses are different from each other, and
therefore in general are non-zero; this means that in general
there are right-handed neutrinos as well as left-handed ones. The
right-handed neutrinos, however, interact with matter only very
weakly. That is why they have not yet been observed.

Now how does the interference work? The wave packets which
represent electron and muon neutrinos are constituted by the sum
of the wave packets which represent mass eigenstates. These wave
packets will travel with different speeds, hence, even though they
may be centered at the same point in space at the point of
production, there will be a seperation between the packets at the
point of detection. In order for interference to be observed,
however, the spatial separation between them must not be too
large. The waves will then interfere at the detector and we can
observe the  phase difference. For more detailed discussion see
Refs.\cite{r8,sto}.

The model which explains this phenomenon realistically is the radial
motion analysis. We begin the investigation, however, by studying
azimuthal motion. Although not realistic, this has the virtue of
simplicity; and it has also been considered in Ref.\cite{r9}, though in a
different formalism. Our method of procedure will be to  write down the
Dirac equation and  find phases corresponding to  mass eigenstates, then
finally calculate phase differences.

\subsection{The Azimuthal Motion}

The Hamiltonian for the azimuthal motion, that is,
$ \vec{p} = (p_r, p_\theta , p_\vp) = (0,0,p) $
could be written from eqn.(\ref{eq15}) as
 \ba
   H =  f \a^3 p_\vp c + f \beta mc^2
          + \f{3}{2} \hb c f A^0 \g_5
          + \f{3}{2} \hb c f \vec{\Sigma}.\vec{A} \; .
 \ea
This Hamiltonian couples the positive energy states to the negative energy
states because of $ \a^3 $ and $ \g_5 $ :
 \ba
    H \psi = E \psi
 \ea
where $ \psi $ is a four-component spinor whose first two
components correspond to positive energy and second two to
negative energy states. Firstly we decouple these states (or block
diagonalize the Hamiltonian) by writing
 \ba
      \rm{det}(H-E)=0
 \ea
where $ E = \left (
                 \begin{array}{cc}
                                   E_+  &   0  \\
                                   0     & E_-
                 \end{array}
          \right ) $.
Therefore the upper block $ E_+ $ and lower block $ E_- $ satisfy
 \ba
    E_+ &=& fpc + \f{fm^2 c^3 }{2p} + \f{3}{2} \hb c f ( -A^0 \s^3 +
           \vec{\s} . \vec{A} )  \\
    E_- &=& - fpc - \f{fm^2 c^3 }{2p} + \f{3}{2} \hb c f (- A^0 \s^3 +
           \vec{\s} . \vec{A} ) \; .
 \ea
Here we assume that the torsional terms are small compared with
the momentum and mass terms\footnote{ Numerical values are given
at the end of next subsection }. With a unitary transformation
 \ba
     \psi = U \xi
 \ea
where U is a unitary matrix and $   \xi =  \left (
                 \begin{array}{c}
                                   \xi_+ \\
                                   \xi_-
                 \end{array}
          \right ) $, our equation system
decouples into two equations
 \ba
    E_+ \xi_+ &=& i \hb \f{\p }{\p t} \xi_+  \label{eq16} \\
    E_- \xi_- &=& i \hb \f{\p }{\p t} \xi_- \; .
 \ea
We will be interested in only the positive energy states for
simplicity. Similar calculations may be done for the negative
energy states. At the end these states will correspond to the mass
eigenstates whose linear superpositions will define, for example,
electron and muon neutrinos as given by (\ref{mix}) and
(\ref{mix1}). Now eqn.(\ref{eq16}) can be cast easily as two
separate equations by diagonalizing, one for spin up, one for spin
down\footnote{ Spin operator is chosen as $ \vec{\Sigma} $.
Besides, it is only in the MASSLESS limit that the neutrino
helicity is $ \pm 1 $, so the state we are discussing is actually
allowed, since we are discussing MASSIVE Dirac neutrinos. };
 \ba
   \{ fpc + \f{fm^2 c^3}{2p} + \f{3}{2} \hb c f A  \}
 \xi^{\ua}_+ &=& i \hb \f{\p }{\p t} \xi^{\ua}_+  \\
 \{ fpc + \f{ f m^2 c^3}{2p} - \f{3}{2} \hb c f A  \}
 \xi^{\da}_+ &=& i \hb \f{\p }{\p t} \xi^{\da}_+ \; ,
 \ea
whose time evolutions are given by
 \ba
   \xi^{\ua}_+(t) &=& e^{-i \Phi^{\ua}(t) } \xi^{\ua}_+(0)  \\
   \xi^{\da}_+(t) &=& e^{-i \Phi^{\da}}(t) \xi^{\da}_+(0)
 \ea
where $ A = \sqrt{( A^0 - A^3 )^2 +{(A^1)}^2 +{(A^2)}^2 }$. The
phase of the spin up state is given by
 \ba
   \Phi^{\ua}(t) = \f{1}{\hb} \int{ \{ fpc + \f{f m^2 c^3 }{2p}
                   + \f{3}{2} \hb c f A  \} dt } \; .
 \ea
For ultrarelativistic neutrinos $ pc \simeq E $ and $ cdt \simeq R
d\varphi \; , $ so the above becomes
 \ba
   \Phi^{\ua} = \f{1}{\hb}  \{ \f{fER}{c} \Delta \varphi
   + \f{f m^2 c^3 R}{2E}\Delta \varphi +
    \f{3}{2} \hb f A R \Delta \varphi \} \; . \label{eq17}
 \ea
Similarly the phase of the  spin down state is
 \ba
 \Phi^{\da} = \f{1}{\hb}  \{ \f{fER}{c} \Delta \varphi
   + \f{f m^2 c^3 R}{2E}\Delta \varphi -
   \f{3}{2} \hb f A R \Delta \varphi \} \; .  \label{eq18}
 \ea
These phases alone do not have an absolute meaning; the quantities
relevant for the interference pattern at the
observation point are discussed below.

Now consider electron and muon neutrinos given by (\ref{mix}) and
(\ref{mix1}) above, where $ \nu_1 $ and $ \nu_2 $ correspond to $
\xi_+ $'s. If electron neutrinos are produced at $ t= 0 $, the
time evolution of $\nu_e $ is
 \ba
   \nu_e (t) &=& \cos\Th \nu_1(t) + \sin\Th \nu_2(t) \nonumber \\
             &=& \cos\Th e^{-i \Phi_1 (t) }\nu_1
             +  \sin\Th e^{-i \Phi_2 (t) }\nu_2 \; . \label{net}
 \ea
The probability of measuring muon neutrino at a later time is given by
 \ba
    P( \nu_e \rightarrow \nu_\mu ) = \sin^22\Th \sin^2\f{ \D \Phi }{2}
      \label{pro}
 \ea
where $ \D \Phi = \Phi_2 - \Phi_1 $. As is seen from (\ref{pro})
the phase difference has an absolute meaning. Now there are four
possibilities.
\begin{itemize}
\item   Both mass eigenstates are spin up, in which case eqn.(\ref{net})
        becomes
        \ba
          \nu_e (t) = \cos\Th e^{-i \Phi^\ua_1 (t) }\nu_1
             +  \sin\Th e^{-i \Phi^\ua_2 (t) }\nu_2 \; . \label{net1}
        \ea
        From (\ref{eq17}) and (\ref{eq18}),
        the phase difference is
        \ba
           \Delta \Phi = \Phi^{\ua}_2 - \Phi^{\ua}_1
         = \f{\Delta m^2 c^3 }{2 \f{E}{f} \hb } R \Delta \vp \; .
         \label{eq19}
         \ea
\item   Both mass eigenstates are spin down. In this case an analogous
        equation to (\ref{net1}) holds and the phase difference
        $ \D \Phi $ is
         \ba
           \Delta \Phi = \Phi^{\da}_2 - \Phi^{\da}_1
         = \f{\Delta m^2 c^3 }{2 \f{E}{f} \hb } R \Delta \vp
           \; . \label{eq20}
         \ea
\item    The first mass eigenstate is spin up, the second spin down,
         then
          \ba
           \Delta \Phi = \Phi^{\da}_2 - \Phi^{\ua}_1
                       = \{ \f{\Delta m^2 c^3 }{2 \f{E}{f} \hb }
             - 3 f A \} R \Delta \vp \; . \label{eq21}
           \ea
\item     The first one is spin down, the second spin up, then
            \ba
           \Delta \Phi = \Phi^{\ua}_2 - \Phi^{\da}_1
                       = \{ \f{\Delta m^2 c^3 }{2 \f{E}{f} \hb }
                         + 3 f A \} R \Delta \vp  \label{eq22}
            \ea
\end{itemize}
where $ \Delta m^2 = m_2^2 - m_1^2 $. From (\ref{eq19}) and
(\ref{eq20}) it is seen that if both mass eigenstates have the
same spin projection there is no contribution to the oscillation
coming from the torsion, but if the mass eigenstates have opposite
spin it is seen from (\ref{eq21}) and (\ref{eq22}) that there is a
contribution to the neutrino oscillation coming from the torsion.

When we do an order of magnitude analysis we see that the
contribution coming from the torsion is really very small compared
with that coming from the mass difference\footnote{ Numerical
values are given at the end of the section 3.2}. Here we also
emphasise that $ \f{E}{f} $ is the locally measured energy
\cite{r6}. If there is no torsion, i.e., $ A^a =0 $, our result
becomes the same as the result of Ref.\cite{r9}, and in this case
there is no gravitational contribution to the phase difference,
since the locally measured energy is $ \f{E}{f} $ rather than $ E
$. We show below that in the case of radial motion the situation
is different.

\subsection{The Radial Motion}

The Hamiltonian for the radial motion, that is, $\vec{p} = (p,0,0) $ may
be written from eqn.(\ref{eq15}) as follows
 \ba
    H =  f^2 \a^1 p_r c + f \beta mc^2
          + \f{3}{2} \hb c f A^0 \g_5
          + \f{3}{2} \hb c f \vec{\Sigma}.\vec{A} \; .
 \ea
Applying the same arguments as in the previous subsection we find
the upper and lower blocks of the Hamiltonian
 \ba
  E_+ &=& f^2 pc + \f{m^2 c^3 }{2p} + \f{3}{2} \hb c f (-A^0 \s^1 +
           \vec{\s}.\vec{ A} ) \\
  E_- &=& -f^2 pc - \f{m^2 c^3 }{2p} - \f{3}{2} \hb c f (-A^0 \s^1 +
           \vec{\s}.\vec{ A} ) \; .
 \ea
As before, we are only interested in the positive energy
states which obey
 \ba
     E_+ \xi_+ &=& i \hb \f{\p }{\p t} \xi_+  . \label{eq23}
 \ea
Again we first diagonalize eqn.(\ref{eq23}) in order to find
eigenvalues of spin up and down states. In this case the phases
appropriate to the spin up and spin down particles are,
respectively,
 \ba
    \Phi^{\ua} &=& \f{1}{\hb} \{ \f{E}{c} \Delta r - \f{2MGE}{c^3}
               \ln{\f{r_B}{r_A}} + \f{m^2c^3}{2E} \Delta r \nonumber \\
              &{ }&  + \f{3}{2} \hb A
    ( \Delta r - \f{MG}{c^2} \ln{\f{r_B}{r_A}}) \} \label{eq24} \\
 \Phi^{\da} &=& \f{1}{\hb} \{ \f{E}{c} \Delta r - \f{2MGE}{c^3}
               \ln{\f{r_B}{r_A}} + \f{m^2c^3}{2E} \Delta r \nonumber \\
              &{ }&  - \f{3}{2} \hb A
    ( \Delta r - \f{MG}{c^2} \ln{\f{r_B}{r_A}}) \}  \label{eq25}
 \ea
where $ A= \sqrt{{(A^0 -A^1)}^2 +{(A^2)}^2 +{(A^3)}^2 } $. Here we
made the assumptions $ pc \simeq E $, $ cdt \simeq dr$, $ f=
(1-\f{2MG}{rc^2} )^{1/2} \simeq 1-\f{MG}{rc^2} $, $ \Delta r = r_B
- r_A $, where A and B refer respectively to the points of
production and detection; hence  $r_A$ is the radius of the sun
and $r_B$ is the distance from the center of the sun to the
surface of the earth.

If we apply the same procedure as in the previous subsection, from
eqn.(\ref{eq24}) and (\ref{eq25}) there are three different phase
differences;
\begin{itemize}
\item  both mass eigenstates have the same spin state
       \ba
          \Delta \Phi = \Phi^{\ua}_2 - \Phi^{\ua}_1
         = \Phi^{\da}_2 - \Phi^{\da}_1
         = \f{\Delta m^2 c^3 }{2 E \hb } \Delta r \label{eq26}
        \ea
\item   the first is spin up, the second spin down
       \ba
       \Delta \Phi = \Phi^{\da}_2 - \Phi^{\ua}_1
                      = \f{\Delta m^2 c^3 }{2 E \hb } \Delta r
       - 3A ( \Delta r - \f{MG}{c^2} \ln{\f{r_B}{r_A}}) \label{eq27}
        \ea
\item  the first is down, the second up
         \ba
   \Delta \Phi = \Phi^{\ua}_2 - \Phi^{\da}_1
               = \f{\Delta m^2 c^3 }{2 E \hb } \Delta r
   + 3 A ( \Delta r - \f{MG}{c^2} \ln{\f{r_B}{r_A}}) \label{eq28}
        \ea
\end{itemize}
where the mass difference is
$ \Delta m^2 = m_2^2 - m_1^2 $.

If an order of magnitude analysis  is done, for vacuum $ \Delta
m^2 c^4 \sim 10^{-10} \;\; eV^2 $, $ E \sim 10 \;\; MeV $, $ r_A
\equiv R_\odot = 7 \times 10^8 \;\;m $, the earth-sun distance $
\Delta r \simeq 1.5 \times 10^{11} \;\; m $ and $ r_B = \Delta r +
r_A \simeq 1.5 \times 10^{11} \; \; m $ and for the sun $
\f{MG}{c^2} \sim 1.5 \;\; km $. As for the magnitude of the vector
$ A^a $, it follows from (\ref{tor}) that
 \ba
    A^a \sim \ell^2  ( \overline{\psi} \psi ) \; .
 \ea
Then $ \Vert A^a \Vert \sim 10^{-62} \; m^{-1} $ and hence the
torsional term in (\ref{eq27}), (\ref{eq28}) is much smaller than
the mass difference term. When we put $ A^a = 0 $ we recover the
same result as Ref.\cite{r9}, but we see that in general there is
a contribution to neutrino oscillation coming from the torsion of
spacetime.

\section{Result}

We have treated the gravitational neutrino oscillation problem in
a different way from previous authors by calculating the
Hamiltonian of a Dirac particle in Riemann-Cartan spacetime and
finding the dynamical phase. We began for simplicity by treating
the (unrealistic) case of azimuthal motion; this gave us the
opportunity to compare with \cite{r9}, in which azimuthal motion
is considered in Riemann spacetime. We have found that there is a
torsional contribution which depends on the spin directions of
mass eigenstates, eqn.(\ref{eq21}) and (\ref{eq22}). If we put the
torsion equal to zero in these equations we recover the same
result as Ref.\cite{r9}. Next we studied the radial motion which
is a more realistic model for our problem. We found here also a
torsional contribution to the neutrino oscillation that is
dependent on the spin polarizations of the mass eigenstates.
Again, when the torsion vanishes we recover exactly the same
result as in ref\cite{r9}.

\bigskip

\noindent {\Large {\bf Acknowledgement}}

One of the authors (M.A.) thanks TUBITAK (Scientific and Technical
Research Council of Turkey) for a grant through BAYG-BDP that
makes his stay at UK possible and the Department of Physics,
University of Kent at Canterbury for hospitality.


\begin{thebibliography}{99}

  \bibitem{r1} W.C.Haxton, B.R.Holstein, {\it Neutrino Physics},
               Am.J.Phys.{\bf 68},15,2000, e-print hep-ph/9905257
  \bibitem{r2} E.K.Akhmedov, {\it Neutrino physics}, Lectures given
               at Trieste Summer School in Particle Physics,
               June 7 - July 9, 1999, e-print hep-ph/0001264
  \bibitem{pon} B.Pontecorvo, {\it Neutrino Experiments and the
                Problem of Conservation of Leptonic Charge},
                Sov.Phys.JETP{\bf 26},984,1968
  \bibitem{tru} A.Trautman, {\it On the Einstein--Cartan Equations II},
                Bull. Acad. Polon. Sci., ser.sci.math.astr.phys. {\bf
                20},503,1972;\\
                A.Trautman, {\it On the Structure of the
                Einstein--Cartan
                Equations}, Symposia Mathematica {\bf 12},139,1973 \\
                T. Dereli, R. W. Tucker, {\it An intrinsic analysis
                of neutrino couplings to gravity}, J. Phys.
                {\bf A15}, 1625, 1982
  \bibitem{r3}  Y.N.Obukhov, {\it On the gravitational moments of a
                Dirac particle}, Acta.Phys.Polon.{\bf B29},1131,1998,
                e-print gr-qc/0001089
  \bibitem{r5} D.V.Ahluwalia, C.Burgard, {\it Gravitationally Induced
               Neutrino Osscillation Phases}, Gen.Rel.{\bf 28},10,1996,
               e-print gr-qc/9603008
  \bibitem{r6} T.Bhattacharya, S.Habib, E.Motolla, {\it Comment on
              "Gravitationally Induced Neutrino-Oscillation Phases"},
                Phys.Rev.{\bf 59},067301,1999, e-print gr-qc/9605074
  \bibitem{r7} D.V.Ahluwalia, C.Burgard, {\it About the Interpretation
               of Gravitationally Induced Neutrino Oscillation Phases},
               e-print gr-qc/9606031
  \bibitem{r8} Y.Grossman, H.J.Lipkin, {\it Flavor oscillations from a
                spatially localized source: A simple general treatment},
                Phys.Rev.D,{\bf55},5,1997
  \bibitem{r9} C.Y.Cardall, G.M.Fuller, {\it Neutrino oscillations in
               curved spacetime: A heuristic treatment}, Phys.Rev.D,
               {\bf 55},12,1997, e-print hep-ph/9610494
  \bibitem{r10}  M.Alimohammadi, A.Shariati, {\it Neutrino oscillation
                in a space-time with torsion}, Mod.Phys.Lett.{\bf A14},
                 267,1999, e-print gr-qc/9808066
 \bibitem{bjo} J.D.Bjorken, S.D.Drell, {\bf Relativistic Quantum
                Mechanics} (McGraw Hill:San Fransisco,1964)
 \bibitem{ryd} P.Singh, L.H.Ryder, {\it Einstein-Cartan-Dirac theory in
                the low-energy limit}, Class.Quan.Grav.,
                {\bf 14},3513,1997
  \bibitem{sto} L.Stodolsky, {\it Matter and Light Wave Interferometry in
                Gravitational Fields}, Gen.Rel.and Grav., {\bf 11},6,1979

\end{thebibliography}
\end{document}